\documentclass{article}

\usepackage[pdftex]{graphicx}

\PassOptionsToPackage{numbers}{natbib}
 \usepackage[preprint]{neurips_2026}

\usepackage[utf8]{inputenc} 
\usepackage[T1]{fontenc}    
\usepackage[hidelinks]{hyperref}       
\usepackage{url}            
\usepackage{booktabs}       
\usepackage{multirow}       
\usepackage{amsfonts}       
\usepackage{nicefrac}       
\usepackage{microtype}      
\usepackage{xcolor}         
\usepackage{float}
\usepackage{changes}        

\title{RoofNet: A Global Dataset for Roof Material Identification from Earth Observation}

\author{Benjamin Tarver$^*$, Noelle Law$^*$, Sasha Getz \& Yuki Miura \\
  Department of Mechanical and Aerospace Engineering, Center for Urban Science and Progress\\
  Tandon School of Engineering, New York University\\
  Brooklyn, NY 11201  \\
\texttt{\{benjamin.tarver, ntl2689, scg9412, yuki.miura\}@nyu.edu} \\
}

\begin{document}

\maketitle

\def\thefootnote{*}\footnotetext{Equal contribution.}\def\thefootnote{\arabic{footnote}}

\begin{abstract}
Building-level exposure data are critical to natural hazard risk modeling, yet most global inventories describe where buildings are located rather than what they are made of. Roof material is a critical but poorly documented attribute for assessing vulnerability to wildfires, wind hazards, urban heat, floods, and earthquakes. 
To address this gap, we introduce RoofNet, 
a global dataset that maps 49,662 georeferenced building instances from 101 countries to 14 key roofing material classes using Earth observation (EO) imagery \textcolor{black}{(redistributed where permitted)} and associated geospatial metadata.\footnote{Data and code are publicly available under the CC BY 4.0 license at the following addresses: https://www.kaggle.com/datasets/doubleblindreview/xbd-roof-images and https://github.com/Climate-Energy-and-Risk-Analytics-Lab/RoofNet.}
RoofNet contributes (1) climatographically and architecturally diverse coverage of roof material labels, (2) a scalable annotation pipeline combining SME-guided manual labeling with vision-language model (VLM)-assisted classification, rule-based validation, and human-in-the-loop verification, and (3) a resource for evaluating subtle, geographically variable material-level identification in EO imagery and its implications for material-aware hazard risk modeling.
Evaluation on a manually labeled hold-out set shows that zero-shot Remote Contrastive Language-Image Pre-Training (RemoteCLIP) struggles with roof material classification, while fine-tuning with RoofNet improves top-1 accuracy from $4.9\%$ to $47.7\%$.
We use RoofNet in an illustrative hazard case study to demonstrate how material-aware exposure data can change vulnerability estimates relative to material-naïve inventories.
RoofNet provides a missing material layer for global building attribute mapping and scalable hazard risk assessment.
\end{abstract}

\section{Introduction}\label{sec:introduction}
As natural disasters become more frequent and severe, there is an urgent need for scalable tools that assess infrastructure vulnerability and support resilience planning. The 2023 Türkiye-Syria earthquakes resulted in an estimated \$34.2 billion in direct damages \citep{liu_what_2024}. Hurricane Helene (2024) and Hurricane Ian (2022) contributed an additional \$78.8 billion and \$112.9 billion in losses, respectively~\citep{noaaofficeforcoastalmanagementHurricaneCosts2025,fema_flood_2025}. Cumulatively, natural hazard events in 2024 alone generated over \$300 billion in global economic impact, underscoring the escalating cost and frequency of disasters~\citep{nationalcentersforenvironmentalinformationnceiUSBilliondollarWeather}. Meanwhile, insurance coverage is declining, and growing populations combined with rapid urbanization are amplifying vulnerability to these events~\citep{nationalcentersforenvironmentalinformationnceiUSBilliondollarWeather,GallagherReNatural2026}. 

Roofing materials are a critical, yet underrepresented component in disaster risk models, directly affecting how structures respond to different hazards. \textcolor{black}{For instance, the Global Earthquake Model (GEM) Foundation documents in their Building Taxonomy~\citep{brzevGEMBuildingTaxonomy2013} how roof material and structural systems generally shape a building's seismic risk profile. \citet{papathoma-kohleWildfireVulnerabilityIndex2022} discuss how roofing material is the variable most strongly associated with a building's vulnerability to wildfires. \citet{ghahremaniHurricaneDamageResidential2024} further discuss how asphalt shingle, sheet metal, and tile roofing all respond differently to hurricane-induced wind stresses. Different roof materials also have different thermal energy storage capacities and albedos, both key variables for studying urban heat island impacts~\citep{departmentofenergyCoolRoofs2025, romanSimulatingEffectsCool2016}.} Accurate identification of roofing materials is also critical for modeling post-disaster reconstruction costs, as recovery operations can impose significant stress on supply chains, resulting in material shortages, cost inflation, and prolonged displacement timelines~\citep{liu_what_2024}. Despite their critical role in estimating natural hazard exposure and reconstruction feasibility, roofing materials remain poorly documented in global datasets~\citep{tingzonFusingVHRPostdisaster2023, guthulaNacalaRoofMaterialDroneImagery2024, mantasRoofSenseMultimodalSemantic2025}. Existing approaches to infrastructure material identification often depend on in-situ surveys or large-scale manual annotation, limiting scalability~\citep{aravena_pelizari_automated_2021,gonzalez_automatic_2020}.

\begin{figure}[h]
\begin{center}
  \includegraphics[width=0.8\linewidth]{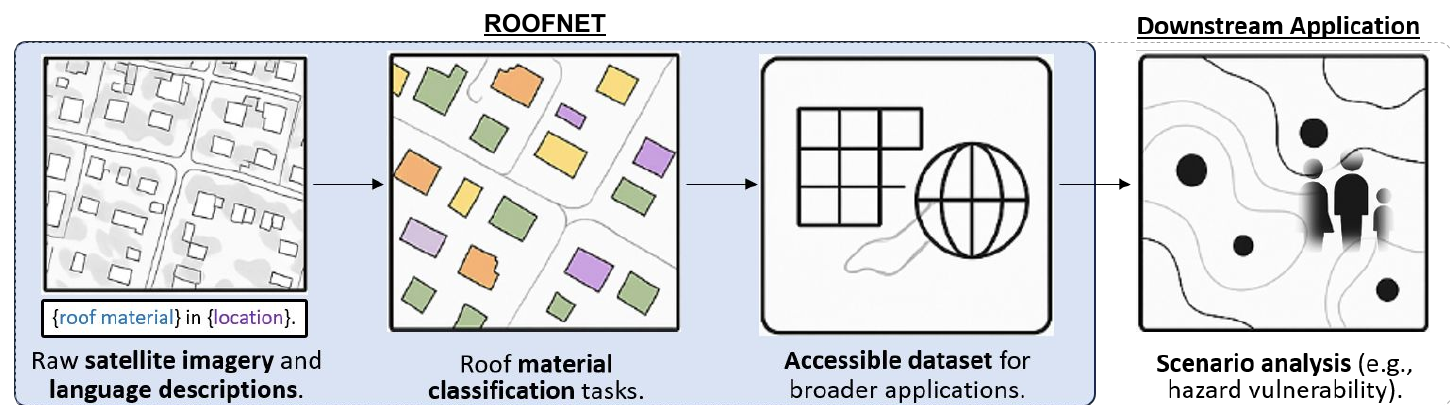}
  \caption{\textbf{Overview of RoofNet and downstream applications}. (1) EO imagery is annotated with prompts describing roof material and location. Materials are labeled using SME validation and VLM-assisted classification. (2) A VLM is trained on a subset of RoofNet to enable \textcolor{black}{semi-}automated classification, with SME validation ensuring reliability of the results. (3) RoofNet \textcolor{black}{metadata} and training information are hosted online to allow for open access to (4) advance risk modeling frameworks and other downstream applications.}
  \label{fig:overview}
  \end{center}
\end{figure}

To address this gap, we introduce RoofNet, \textcolor{black}{a global dataset that uses EO imagery to map georeferenced building instances to curated text descriptions for roof material classification}. RoofNet is designed to enable regional- and global-scale simulations of infrastructure risk, providing the material-specific context that traditional exposure datasets often lack~\citep{guptaXBDDatasetAssessing2019}. Inspired by BRAILS++~\citep{cetinerBRAILS2025, zsarnoczayOpensourceSimulationPlatform2025}, which leverages machine learning \textcolor{black}{(ML)} to generate detailed building inventories for hazard modeling, RoofNet focuses on the problem of \textcolor{black}{roof} material classification. The overview of the dataset and its downstream applications can be seen in Figure \ref{fig:overview}.

RoofNet integrates satellite imagery with open-source geospatial data from platforms such as OpenStreetMap (OSM)~\citep{openstreetmapOpenStreetMap} to label 14 different classes of \textcolor{black}{key} roofing materials for hazard risk analysis, including asphalt shingles, clay tiles, concrete slabs, and metal (corrugated, tiling) materials. The dataset includes 49,662 annotated \textcolor{black}{building instances} sourced from climatically and architecturally diverse regions that span 101 countries across the globe, pairing descriptive material and location labels to support both vision-only and vision-language learning tasks.

By enabling the fine-tuning of VLMs for roof material classification at a global scale, RoofNet establishes a new paradigm for \textcolor{black}{semi-}automated, high-resolution exposure modeling in hazard risk assessment. Unlike prior datasets that are geographically limited or constrained to footprint and height extraction, RoofNet provides the first \textcolor{black}{subject-matter expert} (SME)-curated and vision-language-augmented dataset specifically focused on roof material diversity. The inclusion of metadata such as roof shape, area, solar panel presence, and multi-material indicators further expands its utility across domains. This enables robust model training and cross-regional generalization, while supporting real-world applications such as climate adaptation strategies (e.g., assessing UHI vulnerability), supply chain analysis, and post-disaster insurance modeling. RoofNet offers a set of tools for researchers \textcolor{black}{worldwide in the public and private sectors to perform scalable hazard risk assessments.}

\section{Related Works}\label{sec:related_works}
Building classification and material identification have received growing attention with the advancement of ML and deep learning (DL) techniques \citep{aravena_pelizari_automated_2021,murdoch_residential_2024-1,hoffmannModelFusionBuilding2019,kang_building_2018,yuBuildingInformationModeling2019,ramalingam_building_2025}. Numerous studies~\citep{hoffmannModelFusionBuilding2019,ramalingam_building_2025, deierleinCloudEnabledApplicationFramework2020, zheng_building_2021,aravena_pelizari_automated_2021} have leveraged satellite and street-level imagery, along with open geospatial datasets such as OSM~\citep{openstreetmapOpenStreetMap}, to support infrastructure monitoring and disaster risk analysis at scale.

Several studies have demonstrated the utility of visual learning for extracting urban features. \citet{murdoch_residential_2024-1} used CNNs to classify residential building types from street-view imagery, although the limited availability of street-view data in low-resource settings restricts global applicability~\citep{fry_assessing_2020}. Satellite-based approaches offer broader coverage: \citet{yuBuildingInformationModeling2019} proposed a city-scale BIM framework using EO imagery, and \citet{hoffmannModelFusionBuilding2019} improved classification by fusing aerial and street-level views. DL has also been applied to building instance recognition~\citep{kang_building_2018} and building usage prediction using multimodal inputs~\citep{ramalingam_building_2025, luddeckeImageSegmentationUsing2022}, emphasizing the benefit of integrating semantic and visual cues. \textcolor{black}{\citet{moughnieh_efficient_2025} and \citet{liuGroundingDINOMarrying2024} both proposed transformer-based detection frameworks that are broadly transferable to EO imagery classification tasks.} 

Early efforts at roof classification focused on \textcolor{black}{aerial imagery with narrow geographic distribution}: Table~\ref{tab:benchmark} provides a structured comparison of these efforts and related multimodal datasets. These datasets \citep{solovyevRoofMaterialClassification2020,guthulaNacalaRoofMaterialDroneImagery2024,guptaMappingRefugeeCamps2025,mantasRoofSenseMultimodalSemantic2025,tingzonFusingVHRPostdisaster2023} are constrained by country-specific scopes, modest dataset sizes (2k–22k images), reliance on LiDAR and drone imagery, and narrow taxonomies (4–8 roof classes). More recent multimodal resources leverage global imagery and large-scale pretraining, but do not provide material-specific annotations \citep{guptaXBDDatasetAssessing2019, weiOSW2SAutomaticLabeling2025}.

 \begin{table}[!htpb]
  \caption{High-level comparison of roof classification and related multimodal datasets. \textcolor{black}{Image sources are coded as follows: AE = aerial imagery, LD = LiDAR scans,  DR = drone imagery, and EO = satellite imagery.}}
  \label{tab:benchmark}
  \centering
    \begin{tabular}{p{1.125cm} p{3cm} p{2cm} p{1.375cm} p{1cm} p{3cm} }
\toprule
\textbf{Dataset} & \textbf{Geographic Scope} & \textbf{Image Source} & \textbf{\# Images} & \textbf{Roof Classes} & \textbf{Metadata Included}  \\
\midrule
\citep{solovyevRoofMaterialClassification2020}& St. Lucia, Colombia, Guatemala & AE & 22,553 & 4  \\
\midrule
\citep{guthulaNacalaRoofMaterialDroneImagery2024} & Mozambique & DR & 17,954 & 5  \\
\midrule
\citep{guptaMappingRefugeeCamps2025} & Kenya (Kakuma– Kalobeyei refugee camps) & AE & 2,106 & 7 & Buildings, solar panels, roof materials, sanitation facilities  \\
\midrule
\citep{mantasRoofSenseMultimodalSemantic2025} & The Netherlands & AE + LD & 490 & 8 & Solar panels \\
\midrule
\citep{tingzonFusingVHRPostdisaster2023} & Caribbean (Dominica, St. Lucia) & AE + LD & 8,345 & 5 \\
\midrule
\citep{guptaXBDDatasetAssessing2019} & Global & EO, AE, DR & 165,745 & None &  828,725 image-text pairs\\
\midrule
\citep{weiOSW2SAutomaticLabeling2025} & Global & EO, AE, DR & 163,023 & None & 2,389,973 image-text pairs\\
\midrule
\textbf{RoofNet (Ours)} & Global & EO & 49,662 & 14 & Solar panels, roof shape, location \\
\bottomrule
\end{tabular}%
\end{table}


Recent advances in multimodal representation learning create new opportunities for scalable EO imagery-based classification tasks. For instance, RemoteCLIP~\citep{liuRemoteCLIPVisionLanguage2024}, GeoRSCLIP~\citep{zhangRS5MGeoRSCLIPLarge2024}, and FUSAR-KLIP~\citep{yangFUSARKLIPMultimodalFoundation2026} introduced VLMs tailored for remote sensing, aligning visual and textual semantics to enable zero-shot classification and retrieval. While FUSAR-KLIP is specialized for microwave imagery, GeoRSCLIP and RemoteCLIP both focus on the visible spectrum and exhibit similar performance characteristics, with RemoteCLIP being GeoRSCLIP's more well-established predecessor. 

\textcolor{black}{RoofNet leverages recent advances in DL by using globally available EO satellite and geospatial data sources, including OSM~\citep{openstreetmapOpenStreetMap}, xBD~\citep{guptaXBDDatasetAssessing2019}, and Google Maps~\citep{GoogleMapsPlatform}, to enrich roof imagery with material classifications. Building on the progress of applying CLIP to EO imagery, we fine-tune the RemoteCLIP ViT-L/14 model using labeled training data to embed physically meaningful priors on roofing material into the model’s reasoning process. This reduces dependence on manual annotation to scale our dataset while incorporating SME-informed rules to filter low-confidence predictions into an ``Unknown” class. From this process, we create RoofNet, the first globally distributed dataset for roof material classification in the service of hazard analysis. Its contributions are threefold: (1) climatographically and architecturally diverse sampling across 101 countries, (2) an integrated semi-automated pipeline that leverages VLMs for globally scalable annotation, and (3) a resource for evaluating material-level recognition in EO imagery and its implications for hazard-relevant exposure modeling.}

\section{Methodology}

\subsection{Data Collection and Preprocessing}
RoofNet is globally distributed rather than population-representative, and its sampling strategy balances broad geographic coverage with targeted enrichment of rare material classes. The sampling proceeded in two stages. First, countries were sampled approximately in proportion to population size to provide broad global coverage. Second, to address underrepresentation of rarer materials (e.g., polycarbonate, thatch, glass), we targeted specific cities identified through literature, OSM metadata, and SME input. This dual approach overrepresents some cities relative to population weight, but yields a more materially diverse corpus for model training and evaluation.

\begin{figure}[h!]
  \centering
  \includegraphics[width=0.8\linewidth]{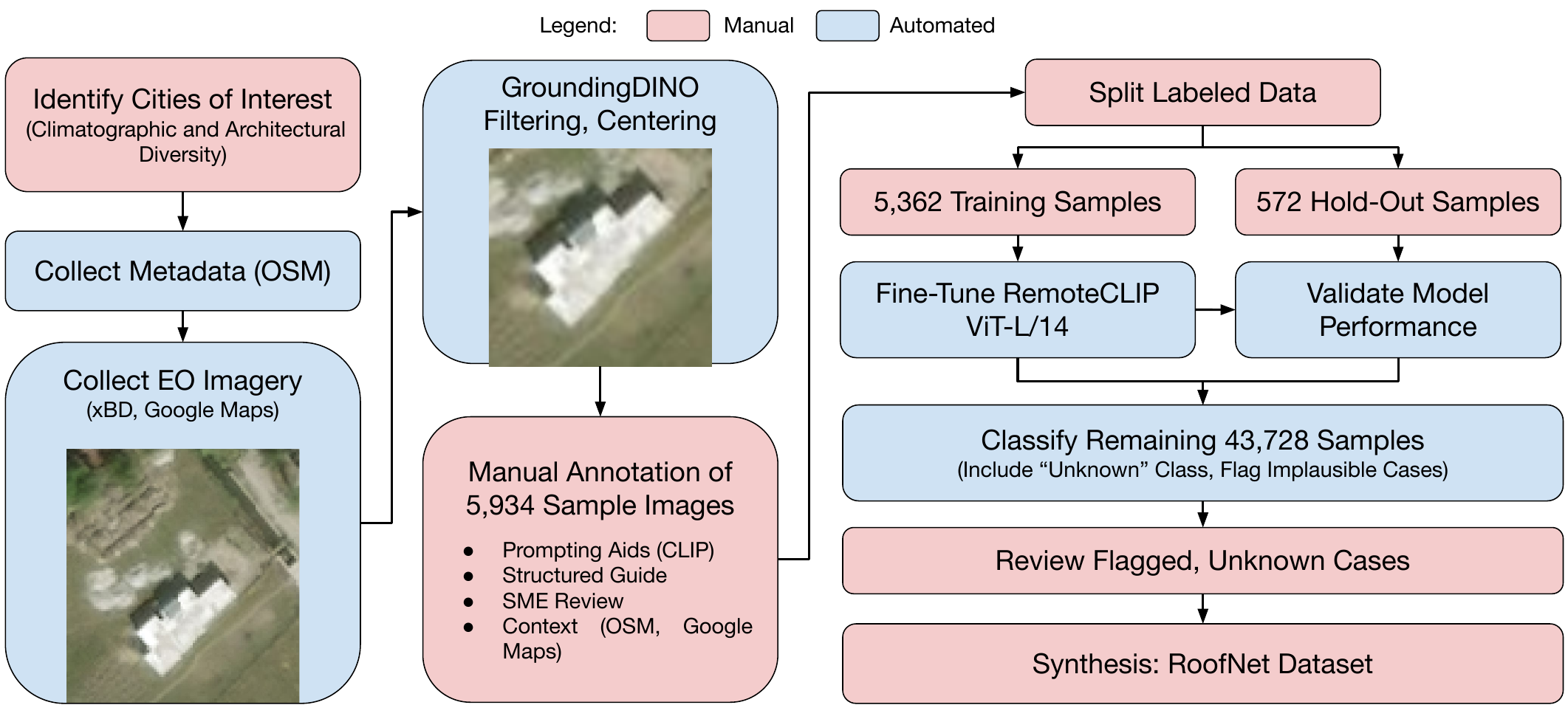}
  \caption{\textbf{RoofNet dataset construction and classification pipeline:} \textcolor{black}{Geographically and architecturally} diverse cities are selected to collect geocoded metadata and representative satellite imagery. Roofs are centered using GroundingDINO~\citep{liuGroundingDINOMarrying2024}, with 5,934 building instances manually verified for fine-tuning and evaluating RemoteCLIP ViT-L/14~\citep{liuRemoteCLIPVisionLanguage2024}. Remaining building instances are classified using the model, followed by rule-based and human-guided validation.}
  \label{fig:curation_overview}
\end{figure}
\begin{figure}[h!]
  \centering
  \includegraphics[width=1\linewidth]{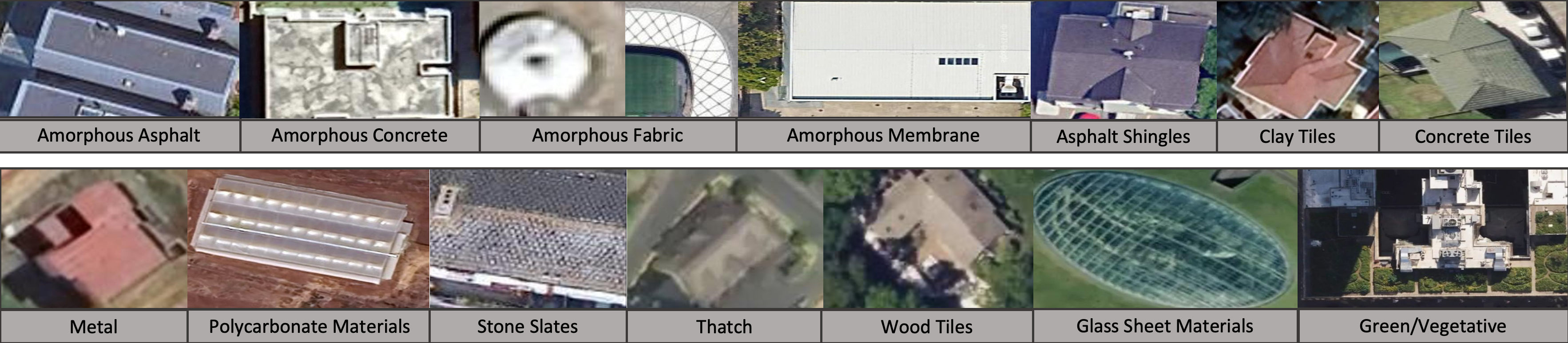}
  \caption{\textbf{RoofNet material classes:}  RoofNet includes 14 roof material classes to support downstream modeling of vulnerability to natural hazards. These classes fall into four general groups: manufactured (clay, concrete, asphalt, wood) tiles, (glass, metal, polycarbonate) sheet materials, synthetic/flat amorphous (asphalt, concrete, fabric, membrane) materials, and traditional/natural (green/vegetative, stone, thatch) materials.}
  \label{fig:classes_visualized}
\end{figure}

Using this approach, approximately 200 buildings per city were geocoded and matched with \textcolor{black}{satellite imagery from Google Maps and xBD~\citep{GoogleMapsPlatform, guptaXBDDatasetAssessing2019}; only imagery with redistribution permission, such as the xBD/Maxar subset, is included in the public release.} We used GroundingDINO~\citep{liuGroundingDINOMarrying2024} with customized prompts to reframe each fetched image around roof structures, improving spatial focus.

Post-processing included filtering out low-resolution or noisy instances and addressing structural ambiguities common in dense urban environments, such as overlapping rooftops or coarse OSM~\citep{openstreetmapOpenStreetMap} polygons representing heterogeneous substructures. Most ambiguous cases were manually re-segmented to isolate individual roof instances, while a minority were retained to preserve diversity and support model generalization under occlusion and spatial complexity.

\subsection{Annotation and Model Training}

To scale classification, we fine-tuned a RemoteCLIP ViT-L/14 model instance on a curated subset of 5,362 building instances \textcolor{black}{(see Figure \ref{fig:curation_overview}), labeled with assistance from a practicing architect}, for 5 epochs~\citep{liuRemoteCLIPVisionLanguage2024}.\footnote{Training required 3 core-hours and testing required 4 core-hours using one CPU core of an Apple M4 Pro, 48\,GB.} Fine-tuning involved an AdamW optimizer with a learning rate of $1\times10^{-5}$
and a batch size of 32~\citep{paszkePyTorchImperativeStyle2019}. \textcolor{black}{We applied image-caption contrastive training to characterize recurrent visual characteristics among roof images of shared material classes. For each material class, we applied descriptive prompts that were iteratively refined for compatibility with RemoteCLIP ViT-L/14 inferential capabilities~\citep{liuRemoteCLIPVisionLanguage2024}. To calculate the similarity between each text embedding and image embedding, we normalized the product of these embeddings by a temperature hyperparameter, set to 0.07 in line with the standard set by}~\citet{radfordLearningTransferableVisual2021} \textcolor{black}{for standard CLIP.} 

\textcolor{black}{To improve model robustness and address sample imbalances, we applied several mitigative strategies in the training and pre-training phases. We sampled image batches in accordance with relative material prevalence using a WeightedRandomSampler() from the torch.utils.data Python package~\citep{paszkePyTorchImperativeStyle2019}. We also implemented label smoothing via the torch.nn.CrossEntropyLoss() function to augment model generalization~\citep{paszkePyTorchImperativeStyle2019}. To reduce prediction noise, we apply a rule-based validator using a region-material lookup table built from literature and SME input. This validator flagged implausible or out-of-distribution predictions (e.g., “Thatch” in Tokyo) for manual inspection to verify predictions based on visual evidence, metadata, and cross-referencing with satellite and street-level imagery. We also enhance in-distribution prediction performance for underrepresented (``long-tail") material classes (i.e., polycarbonate sheets, wood shingles, glass panels) by identifying additional building instances through manual inspection using Google Street View~\citep{google_explore_nodate} and SME input.}



\subsection{RoofNet Metadata}\label{sec:RoofNet_metadata}

To support downstream applications in infrastructure resilience and disaster risk modeling, RoofNet includes auxiliary metadata such as building footprint area, height, roof shape, and number of stories \textcolor{black}{from OSM~\citep{openstreetmapOpenStreetMap}, in addition to solar panel presence and multiple material presence via fine-tuned RemoteCLIP ViT-L/14 prompting~\citep{liuRemoteCLIPVisionLanguage2024}}. These features serve as critical priors for estimating structural vulnerability and hazard exposure~\citep{cetinerBRAILS2025, zsarnoczayOpensourceSimulationPlatform2025,murdoch_residential_2024-1}. For example, footprint area and building height can be used to approximate volumetric mass and potential wind or flood loads, while roof shape is a known determinant of structural performance under wind stress~\citep{internationalcodecouncilinc.2021InternationalBuilding2020,huang_performance-based_2015}. The inclusion of attributes such as solar panel presence and multi-material roofs enables more nuanced assessments of energy infrastructure and construction complexity, which are relevant in post-disaster recovery and climate adaptation planning~\citep{kruitwagen_global_2021,acmsInnovativeBuildingMaterials}. 

\begin{table}[ht]
  \caption{RoofNet metadata overview and scope.}
  \label{tab:metadata_overview}
  \centering
    \begin{tabular}{ p{4cm}p{3cm}  }
    \toprule 
    Metadata     & Coverage (instances) \\
    \midrule 
    \textbf{Solar Panels} & 629      \\
    \textbf{Multiple Materials} & 1,122  \\
    \textbf{Footprint Area (ft$^2$)} & 47,976\\
    \textbf{Roof Shape} & 5,637 \\
    \textbf{Building Height (ft)} & 911 \\
    \textbf{Number of Stories } & 1,234 \\
    \bottomrule
  \end{tabular}
\end{table}

Table \ref{tab:metadata_overview} shows the overall distribution of metadata information for the dataset. The roof material classification further contextualizes roof shape—for example, amorphous asphalt materials such as bitumen are commonly associated with flat roofs, whereas asphalt shingles are typically found on pitched or multi-faceted roof structures.

\subsection{\textcolor{black}{Generalization and Downstream Robustness}}

\textcolor{black}{For downstream disaster response applications, we observed two key source-dependent limitations in direct transfer: image quality and roof material distribution. Some xBD disaster imagery~\citep{guptaXBDDatasetAssessing2019} suffers from visual occlusion and low spatial resolution, creating ambiguous cases where roof material classifications cannot be reliably identified. To address such cases, we introduce an ``Unknown" class to enable the VLM to explicitly abstain from confident prediction. This improves robustness in real-world scenarios, where uncertainty is common during post-disaster response. Furthermore, we observe that the distribution of roof materials and visual features in our Google Maps-derived imagery~\citep{GoogleMapsPlatform} differs significantly from that of open-distribution disaster imagery~\citep{guptaXBDDatasetAssessing2019, VantorOpenData2025}.} In particular, many of RoofNet’s fine-grained distinctions—especially for material classes like Green (Vegetative) Roofs or Polycarbonate Sheet Materials—are difficult to visibly discern in 0.5m resolution xBD imagery, leading to a domain gap in material representation. To address this, we incorporated lower-resolution examples into our model training pipeline to improve generalization in settings where material detail is less pronounced, such as disaster monitoring and damage classification. Tier 1 disaster imagery was extracted from regions including Palu, Indonesia; Southern California; the Midwestern U.S.; Florida; and CDMX, where Tier 1 refers to high-priority events designated by the xBD dataset and selected from the Maxar/DigitalGlobe Open Data Program~\citep{guptaXBDDatasetAssessing2019}.

\section{Dataset Analysis}\label{sec:dataset_analysis}

\subsection{Data Distribution and Coverage}\label{sec:geographic_coverage}
\begin{figure}[h]
  \centering
  \includegraphics[width=1\linewidth]{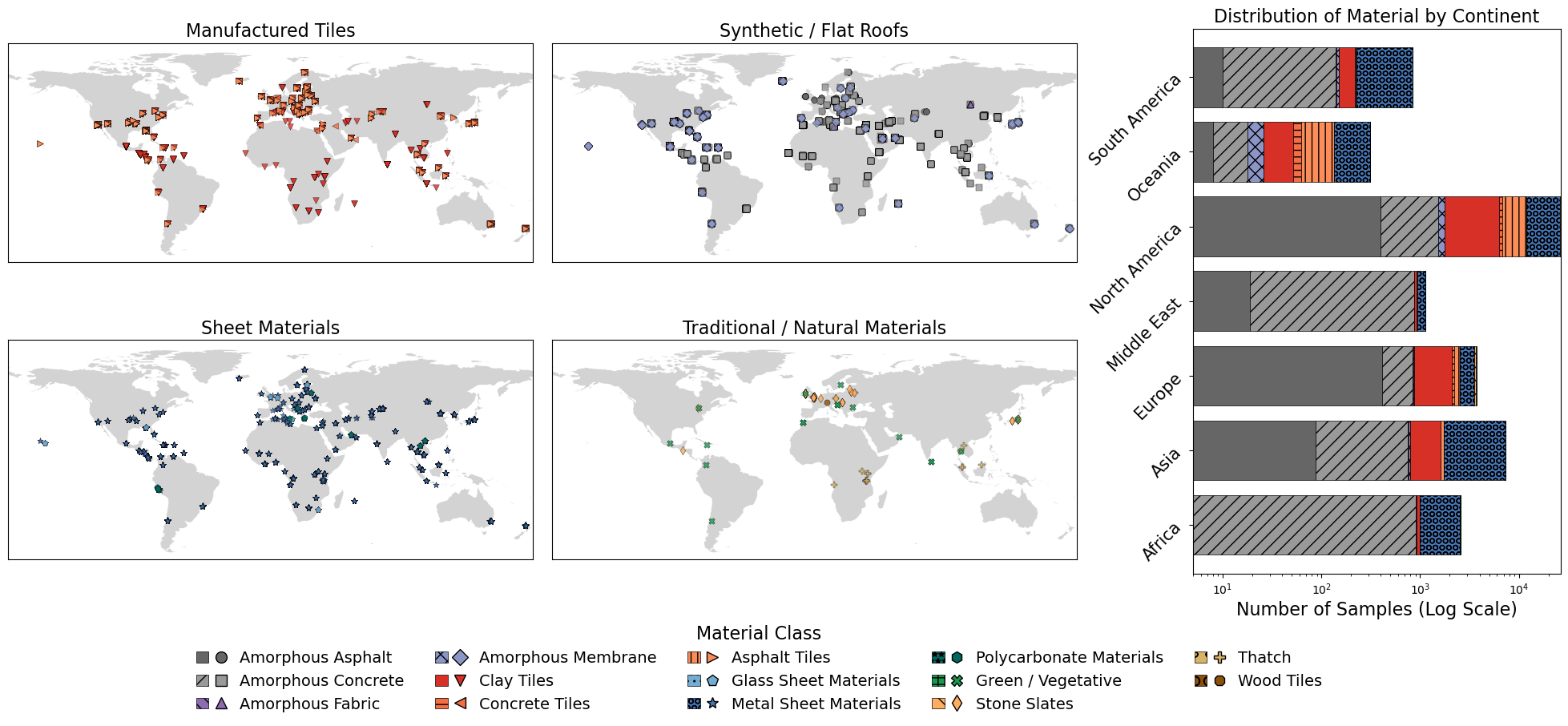}
  \caption{\textbf{Geographic and categorical distribution of roof material classes in RoofNet}. Building instances plotted on the four world maps (left) are colored by material class. These maps illustrate the RoofNet's diverse climatographic distribution. The horizontal stacked bar chart (right) summarizes class prevalence by continent on a log scale, highlighting significant class imbalance and underscoring the need for balancing or augmentation during training.}
  \label{fig:material_distribution_overview}
\end{figure}

\textcolor{black}{Figure~\ref{fig:material_distribution_overview} demonstrates the global distribution of RoofNet data for both frequently and rarely occurring material classes.} A class imbalance can be seen in which metal sheet materials and amorphous concrete/asphalt dominate, while categories such as vegetative/green, thatch, glass sheets, polycarbonate sheets, and fabric are rare. Although infrequent, these long-tail classes capture important real-world use cases, such as polycarbonate roofs in agricultural infrastructure, fabric membranes in stadiums, or glass atria in commercial centers~\citep{acmeplasticsWhyPolycarbonateRight2023}. Their presence highlights the importance of class-aware learning strategies and synthetic augmentation to ensure robust model performance across diverse structural typologies. By combining global reach with both dominant and rare materials, RoofNet provides a \textcolor{black}{diverse} training set that supports downstream applications ranging from hazard vulnerability assessment to reconstruction cost estimation.

RoofNet also captures geographic variability in material appearance, which is critical for generalization. Corrugated metal roofs, for instance, may appear sun-bleached and rusted in equatorial Africa, while the same material in Norway often includes matte, snow-resistant coatings~\citep{sun_study_2024}. Similarly, clay tiles are relatively uniform and terracotta-colored in Mediterranean Europe but weathered and irregular in Latin America~\citep{sun_study_2024}. Such geographically mediated differences illustrate why pairing material classes with contextual metadata is essential for developing global models that scale reliably.

\subsection{Model Evaluation}

\begin{figure}[!h]
  \centering
  \includegraphics[width=1\linewidth]{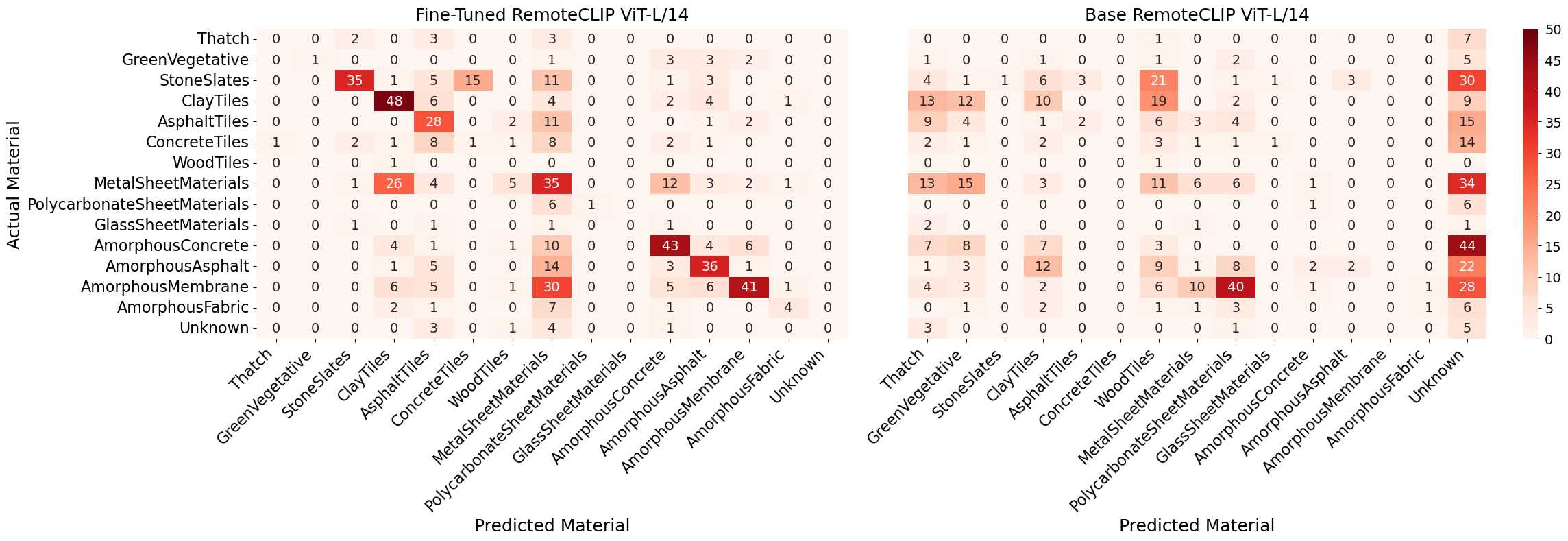}
  \caption{\textbf{Fine-tuned vs. base RemoteCLIP ViT-L/14 evaluation:} Confusion matrices comparing the ability of the base RemoteCLIP ViT-L/14 model (left) and the fine-tuned model (right) to correctly predict the material classes for 572 hold-out roof images.}
  \label{fig:finetune_results_holdout}
\end{figure}

On a 
hold-out set of 572 manually labeled images (14 classes + ``Unknown''), our \textcolor{black}{fine-tuned} RemoteCLIP ViT-L/14 achieves $47.7\%$ top-1 with weighted $\mathrm{P}/\mathrm{R}/\mathrm{F1}=0.540/0.477/0.477$, an order-of-magnitude improvement over zero-shot ($4.9\%$ top-1). Figure \ref{fig:finetune_results_holdout} demonstrates the marked improvement our model fine-tuning yields in RemoteCLIP ViT-L/14's ability to correctly classify roofing materials, demonstrating the model's potential for generating extensible roofing datasets in a semi-automated fashion. 

\textcolor{black}{The base and fine-tuned RemoteCLIP ViT-L/14 models exhibit notable differences in the prediction of "Unknown" and "MetalSheetMaterial" classes.} Specifically, the base model exhibits a much stronger tendency to classify materials as of unknown material type than the fine-tuned model, suggesting that the fine-tuning process incidentally biased the model to favor more certain material predictions. Furthermore, it can be observed that the fine-tuned model overpredicted "MetalSheetMaterials" for most of the material classes and frequently mistook images of metal sheet and amorphous membrane roofs for clay tile and metal sheet roofs, respectively. \textcolor{black}{Such behavior plausibly results from how metal roofing exhibits significant variability on the gray-brown color spectrum based on age- and weather-related factors, which can contaminate classifications of other materials with similar coloration. Table~\ref{tab:group_scores} in Appendix~\ref{app:remoteclip} demonstrates how, for most material groups, group-level performance exceeds material-specific performance, suggesting the model's suitability for capturing aggregated characteristics. For continental-level metrics, see Table~\ref{tab:continent_scores} of Appendix~\ref{app:remoteclip}.} 

We further evaluated our fine-tuned VLM's classification abilities on the full 44,300-sample RoofNet labeled set, including the 572-sample hold-out set and the remaining 43,728 images with semi-automated labeling. On this sample set, we achieve a $40.3\%$ top-1 with weighted $\mathrm{P}/\mathrm{R}/\mathrm{F1}=0.444/0.403/0.389$, while the base RemoteCLIP ViT-L/14 model achieves only a $9.9\%$ top-1. See Figure \ref{fig:finetune_results} and Table \ref{tab:group_scores} in Appendix~\ref{app:remoteclip} for confusion matrices and per-material group metrics.

\subsection{Limitations and Safeguards}\label{sec:limitations}

\textcolor{black}{While RoofNet offers broad geographic coverage and material diversity, there remain issues with effectively sampling diverse metadata on a global scale.} Reliance on OSM~\citep{openstreetmapOpenStreetMap} for metadata (e.g., building footprint, roof shape) restricts RoofNet's utility for downstream applications in regions with scarce metadata coverage. Future work could focus on integrative approaches to metadata collection by leveraging ML models developed to classify roof shape~\citep{huangUrbanBuildingClassification2022, qianDeepRoofRefiner2022, olcerRoofTypeClassification2023} and building height~\citep{kamathGLObalBuildingHeights2024, zhangBuildingHeightExtraction2022, zhaoCombiningICESat2Photons2023} from EO imagery at scale. Furthermore, materials like metal and concrete significantly outnumber rarer classes such as thatch and polycarbonate, and domain shifts in image resolution and quality due to varied data sources may affect model generalization. Future work could address these limitations through targeted data augmentation, inclusion of additional underrepresented regions, and improved resolution harmonization strategies. 
Finally, further model tuning could focus on balancing the inclusion of climatographic priors with building instances that capture intraregional heterogeneity in roof appearance. As discussed by \citet{zhaiAncientVernacularArchitecture2010}, these priors could synthesize climatic, linguistic, and geographic data to illuminate distinct regional architectural traditions.

\section{Case Study Application}\label{sec:case_study}


\subsection{Earthquake Simulation}\label{sec:sim_context}

To demonstrate RoofNet's utility for disaster resilience planning, we present a case study implementation of our data in an earthquake scenario damage simulation. 
We chose to simulate an earthquake due to the availability of relevant building asset data in RoofNet and of globally applicable simulation tools that accept building materials as an input parameter. Conducting a seismic vulnerability assessment allows us to assess the claim that roofing material information, used directly or indirectly, can enhance our understanding of a building's hazard vulnerability. For seismic assessments, it is generally possible to consider a roof's shape, material, structural system, and wall connection as independent variables~\citep{brzevGEMBuildingTaxonomy2013}. However, as \citet{brzevGEMBuildingTaxonomy2013} note, a building's lateral load-resisting system (LLRS) and material are more influential factors in determining its seismic vulnerability. We can thus evaluate the stated claim by comparing the results of an earthquake simulation using roofing material to estimate each building's material and LLRS with a benchmark simulation that is roofing material-naïve.

\subsection{Simulation Methodology}\label{sec:sim_methods}

In this study, we reference the magnitude 7.1 Puebla-Morelos earthquake that struck central Mexico on September 19, 2017~\citep{tena-colungaMexicoCitySeptember2021, tena-colungaPerformanceBuiltEnvironment2020}. More specifically, we simulate the damage effects in CDMX, where over 200 people died and more than 13,000 structures were damaged~\citep{alcantara-ayalaRubbleDisasterRisk2024}. RoofNet contains over 14,000 sample building images from xBD's published pre-disaster imagery for CDMX~\citep{guptaXBDDatasetAssessing2019}. Combining xBD's damage data with information from the GEM Foundation's Global Earthquake Impact Database (GEID), we identify 113 buildings in our dataset that received at least minor amounts of damage~\citep{guptaXBDDatasetAssessing2019, zadehDevelopmentGlobalEarthquake2026}. To simulate the earthquake, we use the GEM Foundation's OpenQuake Engine, which allows users to simulate historical disasters via its scenario damage assessment tool~\citep{silvaDevelopmentOpenQuakeEngine2014, paganiOpenQuakeEngineOpen2014}. The OpenQuake Engine requires information on each simulated building's structural material and LLRS, which we derive from our roofing metadata using an approach developed in coordination with a practicing structural engineer, as well as information from the census data aggregator IPUMS and the GEID~\citep{silvaDevelopmentOpenQuakeEngine2014, paganiOpenQuakeEngineOpen2014,zadehDevelopmentGlobalEarthquake2026,rugglesIntegratedPublicUse2025}. \textcolor{black}{See Appendix~\ref{app:sim_methods} for} the full set of referenced data sources for the simulation. All results are represented by their means plus or minus two standard deviations from said means, \textcolor{black}{sourced from 100-run Monte Carlo analyses for each simulation mode to propagate uncertainty in ground motion fields and material assignments.}

\subsection{\textcolor{black}{Simulation Results, Discussion, and Limitations}}\label{sec:sim_results}

\begin{figure}[h]
  \centering
  \includegraphics[width=.9\linewidth]{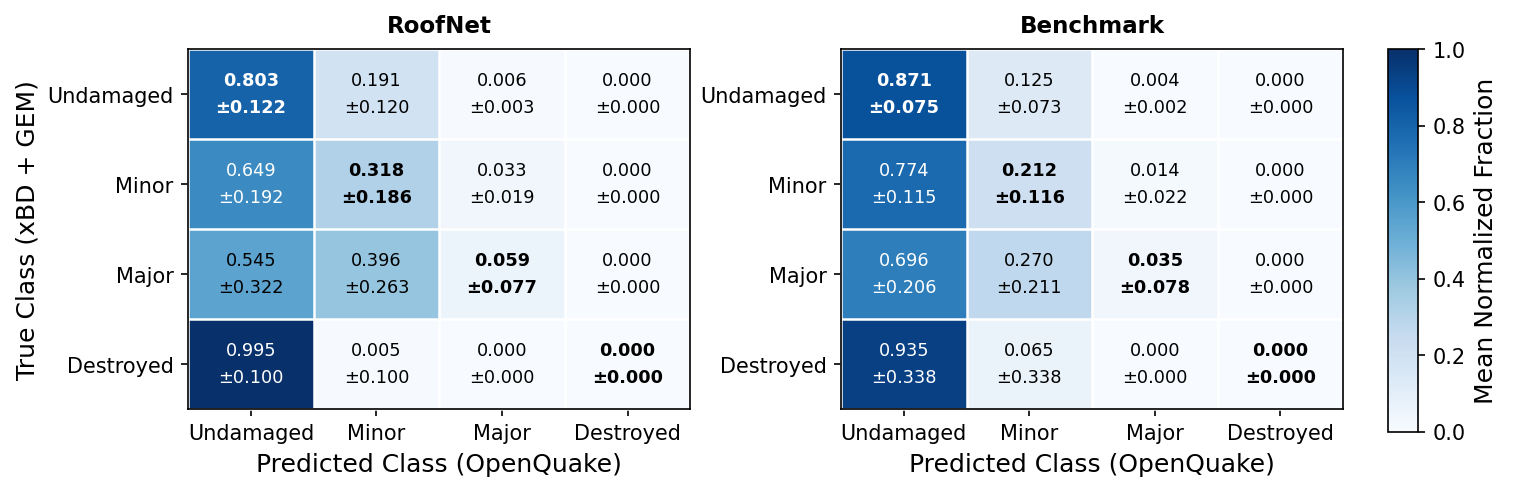}
  \caption{\textbf{Roofing material-aware and material-naïve comparison:} Confusion matrices comparing the normalized benchmark results (right) with roofing material-aware results (left). 14,710 undamaged buildings were sampled for each simulation, in comparison to 91 with minor damage, 20 with major damage, and 2 destroyed.}
  \label{fig:simulation_results}
\end{figure}

Figure~\ref{fig:simulation_results} shows the simulation results comparing the actual damage states recorded by xBD~\citep{guptaXBDDatasetAssessing2019} and the GEID~\citep{zadehDevelopmentGlobalEarthquake2026} with the predicted damage states from the OpenQuake Engine~\citep{paganiOpenQuakeEngineOpen2014,silvaDevelopmentOpenQuakeEngine2014}\textcolor{black}{, illustrating how material-aware exposure assumptions alter predicted damage states}.\footnote{Results computed over 57 core-hours, including <1 core-hour of pre- and post-processing, using 8 CPU cores of an Apple M4 Pro with 48 GB RAM.} \textcolor{black}{Specifically, the incorporation of roofing materials resulted in less conservative damage estimates, yielding a 55\% relative improvement in the simulation's ability to predict damage occurrence for minorly damaged buildings compared to the benchmark (roofing-naïve) approach, in addition to a 49\% improvement for majorly damaged buildings.} A notable consequence of these less conservative estimates is that the roofing material-aware classifications show more variability across seeded runs than in the benchmark case. This is likely a result of the roofing-aware simulations' more aggressive damage scoring overall creating a higher proportion of borderline damage state cases that varied between runs with small perturbations to simulation conditions.

\textcolor{black}{To properly contextualize the simulation results, we discuss key limitations concerning data availability.} First, some damaged buildings may not have been officially recorded as such: While the GEID~\citep{zadehDevelopmentGlobalEarthquake2026} lists over 13,000 buildings damaged by the earthquake, \citet{alcantara-ayalaRubbleDisasterRisk2024} discuss how up to 73,000 properties could have been damaged. The representation problem is echoed by \citet{ainscoeEarthquakeDamageMapped2025}, who discuss how optical imagery, as used by \citet{guptaXBDDatasetAssessing2019}, performs poorly as a tool for remotely assessing earthquake damage to buildings. There are also class distribution discrepancies between the applied data sources that could have downstream impacts: While over 90\% of the residential buildings in the IPUMS data had some form of tile roofing, less than 30\% of roofs in the sampled xBD data match this classification. Furthermore, given the anonymized, aggregated nature of the data published by IPUMS and GEM, it is impossible to validate individual mappings of roofing to building material~\citep{rugglesIntegratedPublicUse2025,yepes-estradaGlobalBuildingExposure2023}. Future work could be supported by the collection of open-access, on-the-ground observations of building and roofing materials in tandem with damage states in sites impacted by natural disasters.

\section{Conclusion}\label{sec:results}
We introduce RoofNet, \textcolor{black}{a dataset of 49,662 georeferenced building instances across 101 countries, designed to address a critical gap in global hazard risk modeling by providing material annotations alongside geospatial metadata}. By combining \textcolor{black}{SME}-reviewed EO imagery with a scalable VLM pipeline, RoofNet advances exposure estimation in hazard-prone regions. Leveraging prompt engineering, targeted data augmentation, and rule-based validation, we achieve accurate classification across 14 roof material types spanning a wide range of climatic and architectural contexts. The dataset is further enriched with metadata such as roof shape, building height, solar panel presence, and footprint area. Beyond material classification, RoofNet holds broader potential for real-world impact, including anticipatory hazard vulnerability assessment, post-disaster supply chain analysis, and reconstruction cost estimation. Indeed, in a set of simulations modeling the 2017 Puebla-Morelos earthquake, we show as much as a 55\% improvement in our ability to correctly backcast damage for buildings by incorporating roofing material knowledge. Looking ahead, future work will explore \textcolor{black}{adaptive extensions that incorporate ML-derived building metadata to assess disaster risks in regions facing well-established and emergent climate hazards.} These enhancements would complement the core dataset, ensuring that RoofNet remains globally inclusive, forward-looking, and a foundation for advancing geospatial AI research in disaster resilience and climate adaptation. 


\section*{Acknowledgements}
The authors wish to acknowledge the statistical office that provided data making the earthquake simulation possible: National Institute of Statistics, Geography, and Informatics, Mexico.

We also wish to thank Mycheal Jonathan-Crafton, Hongjin Zhu, Sarah Wu, and Junyi Li for their valuable contributions and support throughout the development of this work.


\bibliographystyle{unsrtnat}
\bibliography{references}


\appendix

\section{RemoteCLIP Model Performance}\label{app:remoteclip}

\begin{figure}[!htbp]
  \centering
  \includegraphics[width=1\linewidth]{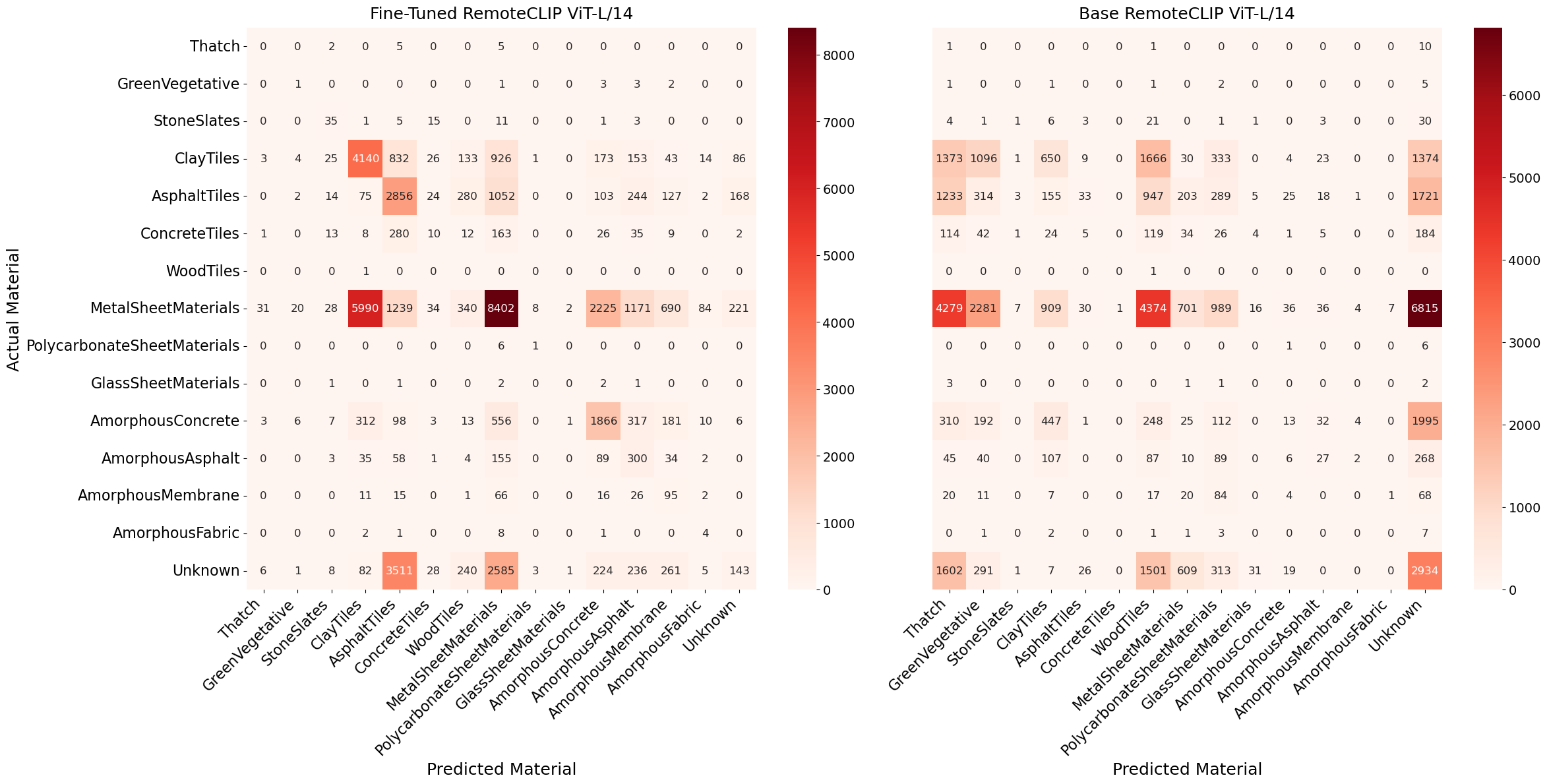}
  \caption{\textbf{Fine-tuned vs. base RemoteCLIP ViT-L/14 evaluation:} Confusion matrices comparing the ability of the base RemoteCLIP ViT-L/14 model (left) and the fine-tuned model (right) to correctly predict the material classes for 44,300 sampled roof images.}
  \label{fig:finetune_results}
\end{figure}

\textcolor{black}{We report results on two evaluation sets: (1) a 572-sample manually labeled hold-out set used for independent evaluation, and (2) the full 44,300-sample RoofNet labeled set produced with assistance from our SME-guided, VLM-assisted annotation framework with rule-based and human-in-the-loop verification. Results on the latter characterize model-label consistency across the released dataset and reveal class- and geography-level failure modes, while the 572-sample hold-out set provides the primary independent accuracy estimate.}

\begin{table}[h!]
  \caption{Per-material group precision, recall, and F1 scoring for base and fine-tuned RemoteCLIP ViT-L/14 models tested on 572 hold-out building instances and the full 44,300 building instances, respectively.}
  \label{tab:group_scores}
  \centering
  \renewcommand{\arraystretch}{1.25}
    \begin{tabular}{ p{3cm}p{4.5cm}p{1.33cm}p{1.33cm}p{1.33cm}  }
    \toprule 
    Model; Instances & Material Group; Instances & Precision & Recall & F1 \\
    \midrule 
    \multirow{4}{*}{Base; 572} & Manufactured Tiles; 134 & 0.294 & 0.112 & 0.162 \\
    & Synthetic/Flat Roofs; 239 & 0.583 & 0.029 & 0.056 \\
    & Sheet Materials; 100 & 0.140 & 0.130 & 0.135 \\
    & Traditional/Natural; 90 & 0.163 & 0.344 & 0.221 \\
    & Unknown; 9 & 0.022 & 0.556 & 0.043 \\
    \midrule 
    \multirow{4}{*}{Fine-Tuned; 572} & Manufactured Tiles; 134 & 0.523 & 0.687 & 0.594 \\
    & Synthetic/Flat Roofs; 239 & 0.770 & 0.632 & 0.694 \\
    & Sheet Materials; 100 & 0.295 & 0.430 & 0.350 \\
    & Traditional/Natural; 90 & 0.704 & 0.422 & 0.528 \\
    & Unknown; 9 & 0.000 & 0.000 & 0.000 \\
    \midrule 
    \multirow{4}{*}{Base; 44,300} & Manufactured Tiles; 12,065 & 0.362 & 0.073 & 0.121 \\
    & Synthetic/Flat Roofs; 4,308 & 0.327 & 0.021 & 0.039 \\
    & Sheet Materials; 20,499 & 0.434 & 0.083 & 0.140 \\
    & Traditional/Natural; 94 & 0.001 & 0.340 & 0.003 \\
    & Unknown; 7,334 & 0.190 & 0.400 & 0.258 \\
    \midrule 
    \multirow{4}{*}{Fine-Tuned; 44,300} & Manufactured Tiles; 12,065 & 0.419 & 0.684 & 0.520 \\
    & Synthetic/Flat Roofs; 4,308 & 0.335 & 0.683 & 0.450 \\
    & Sheet Materials; 20,499 & 0.603 & 0.411 & 0.489 \\
    & Traditional/Natural; 94 & 0.031 & 0.404 & 0.057 \\
    & Unknown; 7,334 & 0.228 & 0.019 & 0.036 \\
    \bottomrule
  \end{tabular}
\end{table}

\begin{table}[h!]
  \caption{Per-continent precision, recall, and F1 scoring for base and fine-tuned RemoteCLIP ViT-L/14 models tested on 572 hold-out instances.}
  \label{tab:continent_scores}
  \centering
  \renewcommand{\arraystretch}{1.25}
    \begin{tabular}{ p{3cm}p{4.5cm}p{1.33cm}p{1.33cm}p{1.33cm}  }
    \toprule 
    Model; Instances & Continent; Instances & Precision & Recall & F1 \\
    \midrule
    \multirow{7}{*}{Base; 572} & Europe; 138 & 0.472     & 0.065  &  0.046\\
    &North America; 311 & 0.062   & 0.556  &  0.111\\
    &Asia; 60 & 0.333 &  0.033   & 0.054\\
    &Africa; 28 & 0.018  &  0.036    & 0.024\\
    &South America; 14 & 0.000   & 0.000  &  0.000\\
    &Middle East; 16 & 0.000  & 0.000 & 0.000\\
    &Oceania; 5 & 0.067  & 0.200  & 0.100\\
    \midrule
    \multirow{7}{*}{Fine-Tuned; 572} & Europe; 138 & 0.760     & 0.587  &   0.651\\
    &North America; 311 & 0.505  &  0.437  &  0.427\\
    &Asia; 60 & 0.399  &   0.433  &   0.409\\
    &Africa; 28 & 0.518  &  0.536  &  0.524\\
    &South America; 14 & 0.243  &  0.143  &  0.148\\
    &Middle East; 16 & 0.812  &  0.750  &   0.780\\
    &Oceania; 5 &  0.200  &  0.200  &  0.200 \\
    \bottomrule
  \end{tabular}
\end{table}

\textcolor{black}{Table \ref{tab:group_scores} reports per-material group basis for both evaluation sets. On the manually labeled hold-out set, fine-tuning improves performance across all known material groups relative to zero-shot RemoteCLIP. Synthetic/flat roofs achieve the strongest group-level performance, while sheet materials remain the most difficult known group, reflecting visual heterogeneity across metal, glass, and polycarbonate roofing materials. The ``Unknown'' class shows the opposite pattern. The base model frequently abstains, whereas the fine-tuned model shifts toward known-material predictions. This suggests that fine-tuning improves material discrimination but reduces conservative abstention behavior.
Results on the full 44,300-sample labeled set reveal how these patterns scale across the released dataset. Performance is strongest for high-frequency groups such as manufactured tiles and sheet materials, while traditional/natural materials remain difficult due to their rarity and visual heterogeneity. The low precision for traditional/natural materials indicates that both base and fine-tuned models produce false positives for long-tail classes, particularly wood tiles and related natural materials. These results highlight the importance of reporting group-level metrics alongside aggregate accuracy.
}

\textcolor{black}{Table~\ref{tab:continent_scores} demonstrates per-continent performance on the manually labeled hold-out set. Performance does not appear to increase monotonically with sample size. The fine-tuned model performs relatively well in Europe and the Middle East, while performance is lower in North America and Asia despite larger or moderate sample counts. This pattern suggests that regional architectural heterogeneity and class composition, rather than sample size alone, influence model performance. These results motivate geography-aware evaluation and future expansion of manually verified labels in regions with diverse material distributions.}

\section{Earthquake Simulation Methodology}\label{app:sim_methods}
This case study evaluates whether roof material information can be integrated into an existing regional seismic risk workflow. We use the OpenQuake Engine scenario damage assessment tool to simulate building damage from the 2017 Puebla-Morelos earthquake in CDMX~\citep{paganiOpenQuakeEngineOpen2014, silvaDevelopmentOpenQuakeEngine2014}. The case study is intended as an illustrative integration experiment, and a roof material is not treated as a direct causal determinant of earthquake damage, but as an additional exposure attribute that can inform probabilistic assumptions about structural material and LLRS.

The OpenQuake Engine requires that the following information be provided to execute a scenario damage assessment: where exposed assets are located, the structural factors dictating said assets' exposure levels (i.e., structural material, LLRS, height, occupancy type), the fragility functions governing said assets' responses to seismic events, and a parameterization of the ground motion fields as they occurred for the studied earthquake~\citep{paganiOpenQuakeEngineOpen2014, silvaDevelopmentOpenQuakeEngine2014}. To represent the earthquake's ground motion fields, we use the ShakeMap Atlas record of ground acceleration values for the 2017 earthquake~\citep{maranoShakeMapAtlas402024}. We use log-log interpolation to compensate for a mismatch between the spectral acceleration values recorded through ShakeMap and those present in the GEM Foundation's record of building typology-specific fragility functions~\citep{maranoShakeMapAtlas402024,martinsDevelopmentFragilityVulnerability2021}. 

\textbf{Exposure attributes.}
For each sampled building, we estimate height, occupancy type, structural material, and LLRS,
which are required parameters for finding a matching fragility function~\citep{martinsDevelopmentFragilityVulnerability2021}. Building height is taken from the UT-GLOBUS dataset, which provides height estimates for individual buildings in CDMX~\citep{kamathGLObalBuildingHeights2024b,kamathGLObalBuildingHeights2024}. The Copernicus program's Global Human Settlement Characteristics (GHS-C) dataset, which reports 10m-scale aggregated estimates of building height, provides a fallback for missing buildings~\citep{pesaresiAdvancesGlobalHuman2024,pesaresiGHSBUILTCR2023AGHS2023}. We also use the GHS-C dataset to distinguish between residential and non-residential land uses, supplemented by global 10m industrial land-use maps to separate industrial and commercial occupancy types~\citep{pesaresiAdvancesGlobalHuman2024,pesaresiGHSBUILTCR2023AGHS2023,yooDatasetMapping10m2025,yooMapping10mIndustrial2025}. 

\textbf{Benchmark assignment.}
To determine the structural material and LLRS for each building, we first reference the GEM Foundation's Global Exposure Model, which provides aggregated sample counts of different building typologies for each of Mexico's federal entities~\citep{yepes-estradaGlobalBuildingExposure2023}. The benchmark simulation relies solely on the distribution of different building typologies in the Exposure Model dataset to pseudo-randomly assign a material and LLRS designation to each building, given its predicted occupancy type. Pseudo-random assignments are made using numpy.random's default\_rng() generator~\citep{harrisArrayProgrammingNumPy2020}, where for each simulation mode, we conduct 100 different runs using integers 0-99 as seeds. 

\textcolor{black}{\textbf{Roof material-aware assignment.}} For the roofing material-aware simulation, we further apply anonymized census data from CDMX on dwelling characteristics, including wall and roof material information, aggregated by IPUMS for the years 2015 and 2020~\citep{rugglesIntegratedPublicUse2025}. This allows us to map between roofing materials and building materials, although this task is complicated by the mismatch between masonry and concrete wall classifications \textcolor{black}{in the Exposure Model~\citep{yepes-estradaGlobalBuildingExposure2023} and in IPUMS~\citep{rugglesIntegratedPublicUse2025}. While the Exposure Model classes specify reinforcement characteristics (e.g., confined masonry, (un)reinforced masonry, reinforced concrete)~\citep{yepes-estradaGlobalBuildingExposure2023}, the IPUMS census data is more generic (i.e., over 99\% of dwellings in CDMX are classified as possessing brick, block, stone, or cement walls)~\citep{rugglesIntegratedPublicUse2025}.} To bridge this gap, we reference on-the-ground observations of the CDMX building stock collected by \citet{tena-colungaPerformanceBuiltEnvironment2020}, who discuss how unreinforced masonry buildings most often have corrugated sheet roofing. We verify the validity of this assumption by conducting additional simulations assuming each masonry structure is either reinforced or unreinforced. For other types of wall materials, we make probabilistic assignments based on the sampled distributions of roof-wall pairings in the IPUMS data~\citep{yepes-estradaGlobalBuildingExposure2023}. 

\textcolor{black}{\textbf{Damage state mapping.}} Since there is a slight mismatch between how damage states are recorded in xBD and by the GEM Foundation, we conservatively map the GEM-derived minor and moderate damage states to xBD's minor damage state. We then score the probabilities as follows: If a certain damage state for a building exceeds the 50\% threshold needed for a majority, then that is recorded as the final damage state. If no damage state exceeds this threshold, then an evenly weighted average of the  OpenQuake damage state probabilities is taken to determine the final damage state.

\subsection{Expanded Simulation Results}

\begin{figure}[H]
  \centering
  \includegraphics[width=1\linewidth]{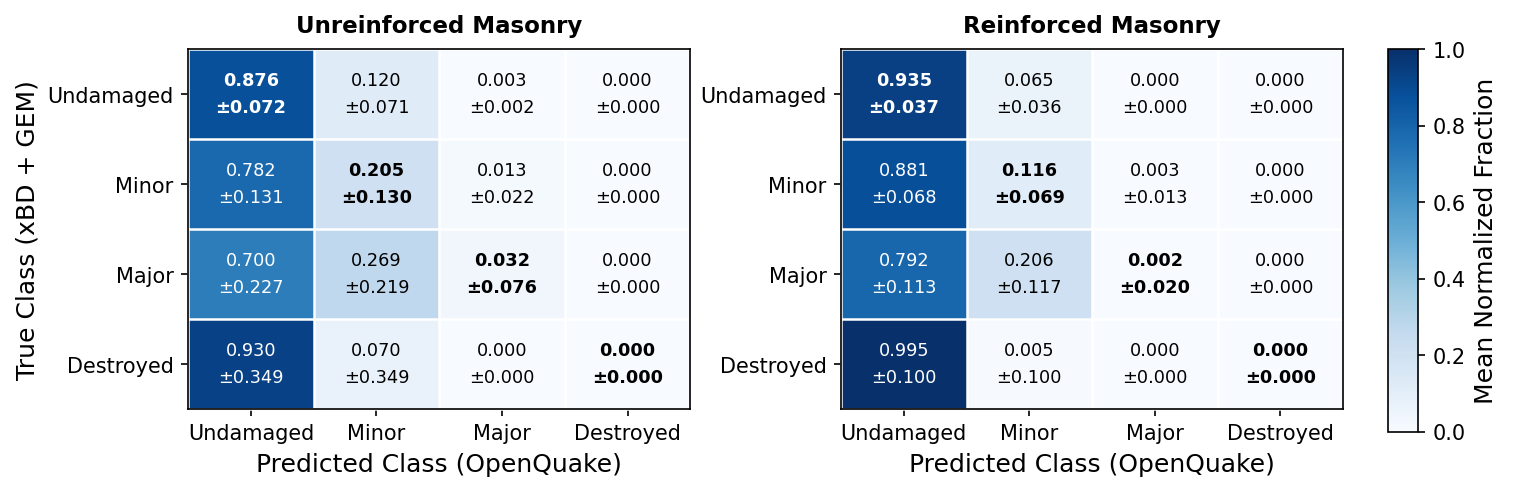}
  \caption{\textbf{Masonry assumption test:} Confusion matrices comparing the results assuming all masonry buildings are unreinforced (left) and assuming all masonry buildings are reinforced (right). All normalized results are represented by their means plus or minus two standard deviations from said means. 14,710 undamaged buildings were sampled for each simulation, in comparison to 91 with minor damage, 20 with major damage, and 2 destroyed.}
  \label{fig:app_simulation_results}
\end{figure}

As shown by Figure~\ref{fig:app_simulation_results}, assuming all masonry or concrete buildings were reinforced led to performance declines in the characterization of buildings that suffered minor and major damage, whereas assuming all such buildings were unreinforced led to similar results to the benchmark case. This result suggests the unreinforced assumption better captures information about the damaged building characteristics.



\end{document}